\documentclass[11pt]{article}
\usepackage{moriond,epsfig}

\def\NPB{{\em Nucl. Phys.} B}
\def\PLB{{\em Phys. Lett.}  B}
\def\PRL{\em Phys. Rev. Lett.}
\def\PRD{{\em Phys. Rev.} D}

\def\PRC{{\em Phys. Rev.} C}

\def\be{\begin{equation}}
\def\ee{\end{equation}}
\def\be{\begin{eqnarray}}
\def\ee{\end{eqnarray}}

\begin{document}
\title{IN MEDIUM MODIFICATIONS TO $\rho$ FROM FINITE PION DENSITY
EFFECTS AND DILEPTON SPECTRUM}

\author{ALEJANDRO AYALA}

\address{Instituto de Ciencias Nucleares, Universidad Nacional 
         Aut\'onoma de M\'exico, Apartado Postal 70-543, 
         M\'exico Distrito Federal 04510, M\'exico.}
\maketitle
\abstracts{
The behavior of the pion dispersion relation in a pion medium
is strongly modified by the introduction of a finite chemical
potential associated to the finite pion number density. Such
behavior is particularly important during the hadronic phase of a
relativistic heavy-ion collision, between chemical and thermal
freeze-out, where the pion number changing processes, driven by the
strong interaction, can be considered to be frozen. We make
use of an effective Lagrangian that explicitly respects chiral
symmetry through the enforcement of the chiral Ward identities. 
The pion dispersion relation is computed through the computation of the 
pion self-energy in a non-perturbative fashion by giving an approximate
solution to the Schwinger-Dyson equation for this self-energy. Given
the strong coupling between $\rho$ vectors and pions, we argue that
the modification of the pion mass due to finite pion 
density effects has to be taken into account self-consistently for the
description of the in-medium modifications of $\rho$'s. We finally
study some possible consequences of finite pion density effects for
the low-mass dilepton spectrum produced in relativistic heavy-ion
collisions.} 

\section{Introduction}

Dileptons are a prime tool to study the evolution of the dense and hot
hadronic region formed in relativistic nuclear collisions. Dilepton
final states are mediated by electromagnetic currents which in turn
are connected to vector mesons. For low invariant masses, the vector
mesons involved are the $\rho$, $\omega$ and $\phi$. Among these,
$\rho$ plays an special role since its lifetime is smaller than the
expected lifetime of the interacting region and thus is able to probe
different stages during the collision of heavy systems.  

Though it is true that a great deal of the features of the measured
low-mass dilepton spectra at SPS
energies~\cite{experiments}$^,$~\cite{Agakichiev} can be 
ascribed to finite baryon density effects~\cite{Rapp}, at
relativistic and even more at ultrarelativistic energies, there is a
large amount of pions produced in the central rapidity
region. Consequently, it is important to include in the calculation of
the dilepton spectrum the proper treatment of the large pion density,
particularly during the hadronic phase of the collision.  

Strictly speaking, pion number is not a conserved
quantity. However, the characteristic time for electromagnetic
and weak pion number-changing reactions, is very large compared to the
lifetime of the system created in relativistic heavy-ion collisions
and therefore, these processes are of no relevance for the propagation
properties of pions within the lifetime of the collision. As for the
case of strong processes, it is by now accepted that they drive pion
number toward chemical freeze-out at a temperature considerably higher
than the thermal freeze-out temperature and therefore, that
from chemical to thermal freeze-out, the pion system evolves with the
pion abundance held
fixed~\cite{Bebie}$^,$~\cite{Braun-Munzinger}. Under these
circumstances, it is possible to ascribe to the pion density a 
chemical potential and consider the pion number as
conserved~\cite{Hung}$^,$~\cite{Chungsik}. 

The description of hadronic degrees of freedom relies on
effective approaches that implement the dynamical symmetries of
QCD. In a series of recent papers~\cite{Ayala}$^,$~\cite{Ayala2} it
has been shown that the linear sigma model can be used as one of such
effective approaches to describe the pion propagation properties
within a pion medium at energies, temperatures and densities small
compared to the sigma mass. The sigma degree of freedom can be
integrated out in a systematic expansion to obtain an effective theory
of like-isospin pions interacting among themselves through an
effective quartic term with coupling $\alpha = 6(m_\pi^2/2f_\pi^2)$,
where $m_\pi$ and $f_\pi$ are the vacuum pion mass and decay constant,
respectively. 

In this work we report on the use of such effective description to
study the interaction of pions with the $\rho$ vector, paying
particular attention to the effects that a finite pion density
introduce on the propagation properties of $\rho$ at finite
temperature. By invoking vector dominance, we also study the
modifications introduced on the production of $e^+$ $e^-$ pairs near
the $\rho$ peak. As a preliminary result, we compare the theoretical
spectrum obtained to SPS data on S + Au collisions at 200 GeV/c
A~\cite{Agakichiev}. A detailed account of this approach can be found
in Ref.~\cite{Ayala3}. 

\section{Effective Lagrangian}

In Refs.~\cite{Ayala} it has been shown that starting from the linear
sigma model Lagrangian, including only meson degrees of freedom and
working in the kinematical regime where the pion momentum, mass and
temperature are small compared to the sigma mass, it is possible to
construct an effective theory that encodes the dynamics of low
energy pion interactions and reproduces the leading order modification
to the pion mass at finite temperature obtained from chiral
perturbation theory~\cite{Gasser}. By {\it gauging} the
theory~\cite{Gale}$^,$~\cite{Pisarski} replacing the
derivative $\partial^\mu$ by the covariant derivative $D^\mu= (\partial^\mu
-ig\rho^\mu)$, where the constant $g$ represents the $\pi$-$\rho$ coupling,
and introducing the mass term and kinetic energy for the $\rho$ field,
the effective Lagrangian for the theory is
\be
   {\mathcal L}\rightarrow {\mathcal L}' =
   \frac{1}{2}(D^\mu\phi)^2
   -\frac{1}{2}m_\pi^2\phi^2 -\frac{\alpha}{4!}
   \phi^4
   +\frac{1}{2}m_\rho^2\rho^\mu\rho_\mu
   -\frac{1}{4}\rho_{\mu\nu}\rho^{\mu\nu}\, .
   \label{effLagprim}
\ee
To introduce a finite chemical potential associated to a conserved pion
number, let us further modify the Lagrangian in
Eq.~(\ref{effLagprim}), writing it in terms of a complex scalar field
and regarding $\phi$ and $\phi^*$ as independent fields
\be
   {\mathcal L}'\rightarrow {\mathcal L}'' =
   (D_\mu\phi)(D^\mu\phi^*)
   -m_\pi^2\phi\phi^* -\frac{\alpha}{4}
   (\phi\phi^*)^2
   +\frac{1}{2}m_\rho^2\rho^\mu\rho_\mu
   -\frac{1}{4}\rho_{\mu\nu}\rho^{\mu\nu}\, .
   \label{effLagdoubleprim}
\ee
It is easy to show that the Lagrangian in Eq.~(\ref{effLagdoubleprim})
leads to a conserved Noether current and to the conserved charge 
\be
   N=i\int d^3x(\phi^*\partial^0\phi - \phi\ \partial^0\phi^* +
   2ig\rho^0\phi^*\phi)\, .
   \label{conscharge}
\ee
that can be identified with the particle number which in turn allows
to introduce a chemical potential $\mu$ conjugate to $N$ which
modifies the grand partition function and translates into a
modification of the Matsubara pion propagator and the $\pi$-$\rho$
vertex which now read as (hereafter capital letters are used to
denote four-vectors whereas lower case letters are used do denote the
components)
\be
   \Delta(i\omega_n,p;\mu)=\frac{1}{-(i\omega_n +\mu)^2 + p^2
   +m_\pi^2}\, ,
   \label{prop}
\ee
and
\be
   \Gamma_{\pi\rho}(P_\mu, {P'}_\mu;\mu)=
   -ig\{[-(i\omega_n + \mu),{\mathbf p}]
   + [-(i{\omega}_{n'} + \mu),{\mathbf p'}]\}\, , 
   \label{vert}
\ee
respectively. 
\section{Self-energies}

At one loop level, the leading contribution to the pion self-energy
$\Pi_0$, including the effects of resummation, corresponds to the {\it
tadpole} approximation~\cite{Hees} and can be obtained from the
solution to the transcendental equation
\be
   \Pi_0=\left(\frac{\alpha T}{4\pi^2}\right)\sqrt{m_\pi^2 + \Pi_0}
   \sum_{n=1}^\infty K_1\left(\frac{n\sqrt{m_\pi^2 + \Pi_0}}{T}\right)
   \frac{\cosh (n\mu /T)}{n}\, .
   \label{solpi0}
\ee
On the other hand, the explicit expression for the one-loop $\rho$
self-energy at finite temperature in the imaginary-time formalism is
given by 
\be
   \Pi^{\mu\nu}=-g^2T\sum_n\int\frac{d^3p}{(2\pi)^3}
   \frac{(2P^\mu-K^\mu)(2P^\nu-K^\nu)}
   {(P^2+\tilde{m}_\pi^2)[(K-P)^2+\tilde{m}_\pi^2]}
   +\delta^{\mu\nu}g^2T\sum_n\int\frac{d^3p}{(2\pi)^3}
   \frac{1}{P^2+\tilde{m}_\pi^2}\, .
   \label{rhoself}
\ee
For a massive vector field, the tensor structure of its self-energy
can be written in terms of the longitudinal $P_L^{\mu\nu}$ and
transverse $P_T^{\mu\nu}$ projection tensors
\be
   \Pi^{\mu\nu}=F(K)P_L^{\mu\nu}+G(K)P_T^{\mu\nu}\, .
   \label{proj}
\ee
The $\rho$ propagator is thus given by 
\be
   -iD^{\mu\nu}=\frac{P_L^{\mu\nu}}{K^2-m_\rho^2-F}+
   \frac{P_T^{\mu\nu}}{K^2-m_\rho^2-G} + 
   \frac{K^\mu K^\nu}{m_\rho^2K^2}.
   \label{proprho}
\ee
In order to identify the coefficients $F$ and $G$, we take ${\mathbf
k}$ along the $z$-axis. Thus, in Minkowski space, their expressions are
\be
   F(K)&=&-\frac{K^2}{k_0k}\Pi^{03}\nonumber\\
   G(K)&=&\Pi^{11}\, ,
   \label{FandG}
\ee
where $\Pi^{03}$ and $\Pi^{11}$ are obtained from Eq.~(\ref{rhoself})
with the analytical continuation 
\be
   i\omega\rightarrow k_0 + i\epsilon\, ,
   \label{analcont}
\ee
that give the retarded functions and that can be performed after carrying
out the sum over the Matsubara frequencies. 

\section{Dilepton rate}

Under the assumption of vector dominance and $\rho$ saturation, the
expression for the rate of dilepton production per
unit space-time volume and unit pair four-momentum is given by~\cite{Weldon}
\be
   \frac{dN}{d^4xd^4k}&=&\frac{1}{3(2\pi)^5}\left(\frac{e^4}{g^2}\right)
   \left(\frac{m_\rho^4}{M^4}\right)
   \Big\{
   \frac{-{\mbox
   {Im}}F}{(M^2-m_\rho^2-{\mbox 
   {Re}}F)^2 + {\mbox
   {Im}}F^2}\left(\frac{1}{e^{\beta\omega_L}-1}\right)\nonumber\\ 
   &+&
   \frac{-2\ {\mbox 
   {Im}}G}{(M^2-m_\rho^2-{\mbox 
   {Re}}G)^2 + {\mbox 
   {Im}}G^2}\left(\frac{1}{e^{\beta\omega_T}-1}\right)
   \Big\}\, .
   \label{ratefin}
\ee   
In order to illustrate the result in Eq.~(\ref{ratefin}), let us take
as a model for the space time history of the collision the Bjorken
model taking as the initial hadron proper-time formation
$\tau_0 = 1$ fm, an initial temperature $T_0=210$ MeV and a final
freeze-out temperature $T_f=130 MeV$. To compute the sound velocity,
we use an equation of state for a hadron gas comprising pions,
nucleons, kaons and $\Delta$(1232). Figure~1 shows the computed
dilepton spectrum for a chemical potential $\mu=100$ MeV and $T=130$
MeV compared to data on S + Au collisions at 200 GeV/c
A~\cite{Agakichiev}. 

\begin{figure}
\begin{center}
\psfig{figure=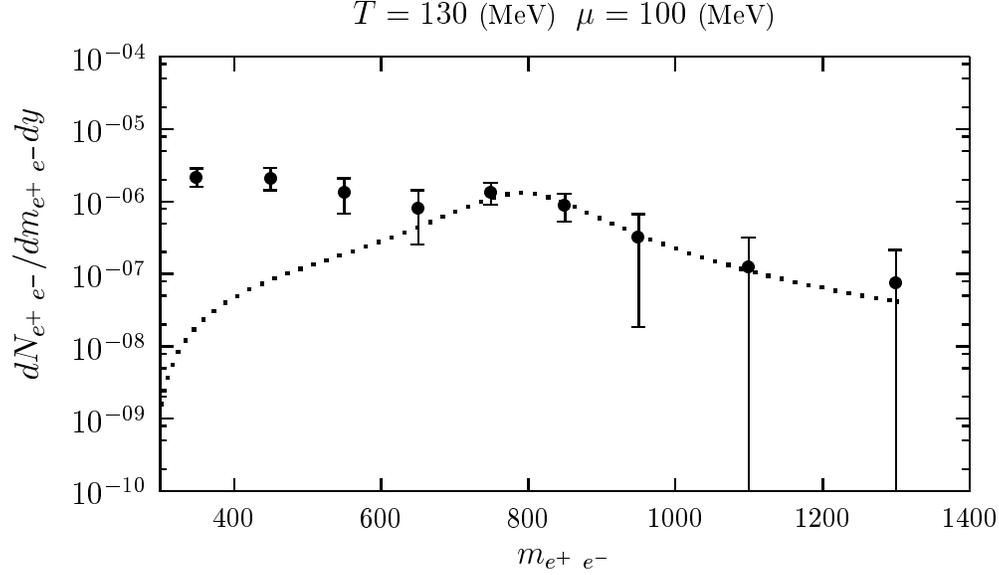,height=3in}
\end{center}
\caption{Dilepton spectrum around the
$\rho$ peak for a chemical potential $\mu=100$ MeV and $T=130$ MeV
compared to data on S + Au collisions at 200 GeV/c
A.\label{fig1}}
\end{figure}

\section{Conclusions}

Under the assumption of VDM and $\rho$ saturation, we have 
computed the dilepton production rate as a function of the pair
invariant mass taking as a model for the space-time history of the
collision the Bjorken model with an equation of state appropriate for a
hadron gas. We have found that the finite pion density produces  
a moderate broadening of the distribution and a moderate increase of
the position of the peak. The finite pion density
also produces a decrease of the distribution at the position of the
peak compared to the $\mu=0$ case.

\section*{Acknowledgments}

Support for this work has been received in part by DGAPA-UNAM under PAPIIT
grant number IN108001 and CONACyT under grant numbers 32279-E and
40025-F.

\end{document}